\documentstyle[11pt,newpasp,twoside]{article}
\begin{document}

\title{Self-Gravitating N-Body Systems out of Equilibrium}
\author{Daniel Huber}
\affil{Observatoire de Gen\`eve, CH-1290 Sauverny, Switzerland}


\begin{abstract}
Real physical systems are often maintained off equilibrium by energy
or matter flows. If these systems are far from equilibrium then the
thermodynamical branch become 
unstable and fluctuations can lead them to other more stable states. 
These new states are often endowed with higher degrees of
organization. In order to explore whether
an energy-flow in combination with self-gravity can lead 
to complex, inhomogeneous structures, like observed in the 
interstellar medium (ISM), we perform $N$-body simulations of
self-gravitating systems subjected to an energy-flow.

Moreover we perform some simple ``gravo-thermal'' N-body experiments 
and compare them with theoretical
results. We find negative specific heat in an energy range as predicted 
by Follana \& Laliena (1999).
\end{abstract}

\section{Introduction}
An energy-flow through a system is related to an entropy-flow. If the 
entropy-flow leaving the system is larger than the entering one, 
the system evacuates its internal, by irreversible processes produced 
entropy to the outer world. Consequently order is created inside the
system, diverging from the classical thermodynamical equilibrium of
a closed system. Far from equilibrium the system is no longer
characterized by an extremum principle, thus losing its
stability. Therefore perturbations can lead to long range order,
through which the system acts as a whole. Such a behavior is well
known in laboratory hydrodynamics and chemistry. The underlying
concepts such as ``dissipative structures'' and ``self-organization''
were extensively studied (see e.g. Nicolis \& Prigogine 1977). But
in spite of their popularity they are until now only little studied in
the context of non-equilibrium structures in self-gravitating
astrophysical systems. Therefore we take up some ideas of these
concepts and build a simple model of an open self-gravitating system. 
With this model we want to check if an energy-flow can maintain a 
self-gravitating system in an statistically stable state, out of 
thermodynamical equilibrium and if gravitation in combination with an 
energy flow can create structures with a higher degree of order.

\section{Dissipative N-body Systems}

Taking into account the highly clumpy nature of the interstellar
medium, we use in our model dissipative particles, representing dense
cloud fragments, to simulate cosmic gas (Pfenniger 1998). 
Moreover, with such a realization we can check some thermodynamic results of
self-gravitating systems.

\section{Model}
In order to prevent gravitationally unbound particles from dissolution, we
confine the particles in a spherical potential well. This prevents 
matter flow. But the system is subjected to an energy flow, maintaining
the system out of equilibrium. This flow
is sustained by energy injection (heating) on large scales and local 
dissipation. The energy injection is due to time and position dependent 
potential perturbations. If the system represents a molecular cloud, then
these potential perturbations can stem from star clusters, clouds or other
high mass objects passing in the vicinity. Indeed such stochastic
encounters must be quite frequent in galactic discs and we assume, 
$1/f_{enc}\ge t_{dyn}$, where $f_{enc}$ is the mean frequency of the
encounters. Thus these encounters can provide a continuing low
frequency energy injection on large scales.
The dissipation is due to  
``inelastic particle encounters''. Therefore we add friction forces to the
equation of motion, depending on the relative particle velocities and 
positions.

If we switch off the heating process and use instead the local
dissipation scheme (``inelastic particle encounters'') a scheme that
dissipates the energy globally, then we can maintain the system during
its evolution nearly in thermodynamic equilibrium. Thus we can perform 
some simple ``thermodynamic experiments'' of self-gravitating systems
with softened potentials.

\section{Results}

Potential perturbations caused by astrophysical objects passing in the 
vicinity of a system can compensate the energy loss due
to dissipation, thus prevent the system from collapsing and
maintain a statistical equilibrium. This statistically steady state
consists of a cold core moving in a hotter halo. The energy-flow leads 
however not to a phase transition to higher
ordered structures.

The gravo-thermal experiments lead to the following results:
A Plummer softened potential yields the same range of negative specific 
heat as found by Follana \& Laliena. The negative specific heat range
is related to a phase transition, separating a high energy homogeneous 
phase from a collapsed phase. As long as cooling processes are at work 
the collapsed phase shows no core-halo structure. After the cooling
has stopped such a structure is formed on the relaxation time scale.

\end{document}